\documentclass[aps,prd,preprint,superscriptaddress,tightenlines,nofootinbib]{revtex4}
\usepackage{graphicx}
\usepackage{dcolumn}
\usepackage{bm}
\usepackage{epsfig}

\def\etal{et al.}

\newcommand{\Psip}{$\psi(2S)$ }

\newcommand{\ee}{$e^{+} e^{-}$ }

\newcommand{\etatogg}{$\eta \to \gamma \gamma $ }
\newcommand{\etatopis}{$\eta \to \pi^{+} \pi^{-} \pi^{0} $ }

\newcommand{\Ks}{$K_{S}^{0} $}

\newcommand{\pio}{$\pi^{0} $}

\newcommand{\chic}{$\chi_c$}

\newcommand{\chicz}{$\chi_{c0}$}
\newcommand{\chico}{$\chi_{c1}$}
\newcommand{\chict}{$\chi_{c2}$}
\newcommand{\pipipiopio}{$\pi^{+}\pi^{-}\pi^{0}\pi^{0}$}
\newcommand{\KKpiopio}{$K^{+}K^{-}\pi^{0}\pi^{0}$}
\newcommand{\pppiopio}{$p\bar{p}\pi^{0}\pi^{0}$}

\newcommand{\KKetapio}{$K^{+}K^{-}\eta\pi^{0}$}

\newcommand{\KpiKopio}{$K^{+}\pi^{-}K^{0}\pi^{0}$}
\newcommand{\KpmpimpKopio}{$K^{\pm}\pi^{\mp}K^{0}\pi^{0}$}
\newcommand{\KpmpimpKspio}{$K^{\pm}\pi^{\mp}K_{S}^{0}\pi^{0}$}

\newcommand{\fonpiopio}{$f_{0}(980)\pi^{0}\pi^{0}$}
\newcommand{\ftpiopio}{$f_{2}(1270)\pi^{0}\pi^{0}$}
\newcommand{\rhopmpimppio}{$\rho^{\pm}\pi^{\mp}\pi^{0}$}
\newcommand{\rhoppimpio}{$\rho^{+}\pi^{-}\pi^{0}$}
\newcommand{\rhompippio}{$\rho^{-}\pi^{+}\pi^{0}$}
\newcommand{\rhozpippim}{$\rho^{0}\pi^{+}\pi^{-}$}
\newcommand{\fonpippim}{$f_{0}(980)\pi^{+}\pi^{-}$}

\newcommand{\KstarpmKmppio}{$K^{*\pm}K^{\mp}\pi^{0}$}
\newcommand{\KstarpmtopipmKo}{$K^{*\pm} \to \pi^{\pm} K^{0}$}

\newcommand{\fonKpKm}{$f_{0}(980)K^{+}K^{-}$}

\newcommand{\KstaroKopio}{$K^{*0}K^{0}\pi^{0}$}
\newcommand{\KstarotoKpmpimp}{$K^{*0} \to K^{\pm}\pi^{\mp}$}

\newcommand{\rhopmKmpKo}{$\rho^{\pm}K^{\mp}K^{0}$}

\newcommand{\KstaroKpmpimp}{$K^{*0}K^{\pm}\pi^{\mp}$}
\newcommand{\KstarotoKopio}{$K^{*0} \to K^{0}\pi^{0}$}
\newcommand{\KstarpmpimpKo}{$K^{*\pm}\pi^{\mp}K^{0}$}
\newcommand{\KstarpmtoKpmpio}{$K^{*\pm} \to K^{\pm}\pi^{0}$}

\newcommand{\gevcsq}{\mbox{\rm GeV/$c^2$}}

\begin{document}

\preprint{CLNS 07/2007}       
\preprint{CLEO 07-12}         

\title{First Observation of Exclusive {\boldmath $\chi_{cJ} $} Decays to Two Charged
and Two Neutral Hadrons}
%

\author{Q.~He}
\author{J.~Insler}
\author{H.~Muramatsu}
\author{C.~S.~Park}
\author{E.~H.~Thorndike}
\author{F.~Yang}
\affiliation{University of Rochester, Rochester, New York 14627, USA}
\author{M.~Artuso}
\author{S.~Blusk}
\author{S.~Khalil}
\author{J.~Li}
\author{N.~Menaa}
\author{R.~Mountain}
\author{S.~Nisar}
\author{K.~Randrianarivony}
\author{R.~Sia}
\author{N.~Sultana}
\author{T.~Skwarnicki}
\author{S.~Stone}
\author{J.~C.~Wang}
\author{L.~M.~Zhang}
\affiliation{Syracuse University, Syracuse, New York 13244, USA}
% ...
\collaboration{CLEO Collaboration} 
\noaffiliation

\date{June 6, 2008}

\begin{abstract} 
We study exclusive $\chi_{c0,1,2} $ decays to 
four-hadron final states involving two charged and two neutral mesons:
\pipipiopio, \KKpiopio, \pppiopio, \KKetapio, and \KpmpimpKopio. The
$\chi_{c} $ states are produced in radiative decays of 3.08 million \Psip
resonance decays and observed in the CLEO detector. We also measure the largest
substructure contributions to the modes \pipipiopio\ and \KpmpimpKopio. 
\end{abstract}

\maketitle


Exclusive charmonium decays have been a subject of interest for
decades as they are an excellent laboratory for studying
quark-gluon dynamics at relatively low energies. 
However, current measurements in the $P$-wave \chic\ sector
are sparse~\cite{PDG}.
Although these states are not directly produced in \ee
collisions, 
they are copiously produced in the 
radiative decays \Psip $\to \gamma \chi_{c} $, each of which has a 
branching ratio of around 9\%~\cite{psi2stogamchicj}. Recent data
taken by the CLEO detector at the Cornell Electron Storage Ring to study 
$e^+e^-$ annihilations with a center of mass energy 
corresponding to the \Psip mass allow for a detailed study of \chic\
decays.

Past research indicates that
the Color Octet Mechanism (COM) plays a role in the decay of these
$P$-wave charmonia states~\cite{1,2,3,4,5}.
In order to build a comprehensive understanding about the $P$-wave
dynamics, both theoretical predictions employing the COM
and new precise experimental measurements for \chic\ many-body final
states are required. Furthermore, decays of $\chi_{c} $, in
particular $\chi_{c0,2}$, may provide a window on glueball
dynamics~\cite{glueballs}.

This analysis follows the general method of our earlier work on 
$\eta,\eta^{\prime} $~\cite{etaeta analysis} and
three-body~\cite{3 body analysis} decays of \chic, extending it to 
higher multiplicity states.  
The four-body exclusive \chic\ decay modes studied in this article
contain two neutral and two charged hadrons in the final state and
are being measured for the first time. 
We also take a first look at the gross features of the rich
substructure in these decay modes.

The data used in this analysis consist of 2.74 pb$^{-1}$ and 2.89
pb$^{-1}$, a total of (3.08$\pm $0.09) $\times $10$^6 $ \Psip decays, taken
with the CLEO~III~\cite{cleo} and  
CLEO-c~\cite{cleoc} detector configurations, respectively. 
Both
detector configurations provide 93\% solid angle coverage, and have
common components of particle identification which are critical to
this analysis: the main drift chamber, the Ring Imaging Cherenkov
Detector (RICH), and the CsI calorimeter (CC). We distinguish two
regions of the CC in polar angle: the barrel ($|\cos(\theta)| < $ 0.81)
 and the endcap ($|\cos(\theta)| \geq $ 0.81).
The CC
detects photons with an energy resolution of 2.2\% (5\%) for photons
with energy of 1~GeV (100~MeV), making it possible to resolve the 
three \chic\ states.

The event reconstruction and final state selection criteria proceed along
the lines of our recent analyses on this 
subject~\cite{etaeta analysis, 3 body analysis}.
We reconstruct the following \chic\ decay modes:
$K^+ K^- \eta \pi^0$; $K^{\pm} \pi^{\mp}K_{S}^{0} $\pio; and
$h^+ h^- $\pio \pio, where $h=\pi, K, p$. 
In addition, we reconstruct the transition
photon from $\psi(2S)$, thus detecting the entire event.

Each charged particle in the event is required to pass standard
criteria for track quality and geometric acceptance. 
We also require the number of such
tracks to be either two or four depending on the final state. 
We demand all tracks come from the beam spot with a
momentum-dependent cut on the impact parameter. This cut is less restrictive
for low-momentum tracks, for which the resolution is poorer.
We combine 
ionization loss in the drift chamber ($dE/dx $) and RICH information
to discriminate between $p, K {\rm, and }~  \pi $ using $\chi^2$ criteria 
discussed elsewhere~\cite{our paper}. 
Additional requirements suppress charged lepton QED backgrounds.
We reject electron candidates as follows: for all tracks, we compute
the ratio of CC energy to track momentum, $E_{{\rm CC}}/p$, and the
difference between the measured $dE/dx$~and the expected $dE/dx$~for
the electron hypothesis, normalized to its standard deviation, $\sigma_{e}$.
We reject tracks with both $0.92~<~E_{{\rm CC}}/p~<~1.05$~ and $|\sigma_{e}|
<~3$.
Particles that penetrate
more than five nuclear interaction lengths of the muon detectors are rejected.
The particle identification together with the lepton veto
criteria are found to be more than 97\% efficient for all the modes.

We define photon showers as those having a lateral profile in the CC
consistent with a photon and possessing at least 30 MeV
of energy. We require photon candidates found in the endcap CC region
that are used in \pio\ and $\eta $ reconstruction
to possess more than 50~MeV of energy.
We reconstruct $\pi^0 \to \gamma \gamma $ and $\eta \to \gamma \gamma
$ candidates using a pair of photons that are kinematically fit to the
nominal \pio\ or $\eta $ mass using the 
event's charged tracks 
to define
the origin of the photon trajectories.
A cut is placed on the mass fit of 
$\chi^2 < 10 $ (for one degree of freedom). 
Each \pio\ and $\eta $ photon daughter is forbidden 
to be part of any other final state particle.
We also reconstruct 
$\eta \to \pi^+ \pi^- \pi^0 $ decay by combining a pair of charged
pion candidates with a \pio\ and 
doing a mass-constrained fit
to the nominal $\eta $
mass with the requirement
$\chi^2  < 10$ for one degree of freedom.    
We reconstruct the \Ks\ by constraining a 
pair of oppositely charged pions to come from 
a common vertex. 
We
require that the reconstructed invariant mass of the two pions be within 10~MeV
($\approx 3.2$ standard deviations) of the nominal \Ks\ mass.  

We combine any unused photon with the four hadrons to reconstruct the
complete event which we then constrain to the
four-momentum of the $\psi(2S)$, using the nominal mass
of the $\psi(2S)$ and taking into account the crossing
angle between the $e^{+} $ and $e^{-} $ beams ($\approx$ 4 mrad). 
We demand  
$\chi^2 < 25 $ (for four degrees of freedom) for this constraint.
This requirement strongly rejects background, and the
fitting procedure greatly improves the mass resolution of the
$\chi_c$. 
In $\leq 10$\% of the events (the $\gamma $\pipipiopio\ state is worst),
multiple possible pairings of photons lead to multiple $\chi_c$
candidates. We choose the one with the smallest $\chi^2 $.



We study the efficiencies and resolutions of the final states by
generating signal events using a {\sc Geant}-based~\cite{Geant} detector
simulation. Over 0.7 million events were generated for the two detector
configurations, three \chic\ mesons, and five final states. 
Events were generated in accordance with an electric dipole (E1) transition
production cross section of $ 1 + \lambda \cos^2{\theta}$,
where $\theta $ is the radiated photon angle relative
to the positron beam direction, and $\lambda=1,-1/3,+1/13$ for
$J=0,1,2$ particles, respectively.

The efficiencies averaged over the CLEO~III and CLEO-c datasets
(weighted by the number of \Psip events) are listed in 
Table~\ref{Table:table_YandE_paper_NR} for each final state. The efficiency
includes the \etatogg, \etatopis and \Ks\ $\to \pi^+ \pi^- $ 
branching ratios~\cite{PDG}. 
The invariant mass distributions of the final state hadrons were
fitted to three signal shape functions corresponding to each of the
three $\chi_{c}$ states and an additional constant background
function. Each signal shape function consisted of a Breit-Wigner
convolved with a double Gaussian resolution function. The widths of the
Breit-Wigner functions are 
fixed to the intrinsic widths of the $\chi_{c}$ states:
$\Gamma_{\chi_{c0}} = 10.4~{\rm MeV} $, $\Gamma_{\chi_{c1}} = 0.89~{\rm MeV} $, and $\Gamma_{\chi_{c2}} = 2.06~{\rm MeV}$~\cite{PDG}.
The detector resolution is  
obtained from simulation by 
a fit to the difference between the generated and
reconstructed invariant mass of the $\chi_c$ products, 
mode-wise and separately for each $\chi_{c}$.
The detector resolution, ranging from $4.5$ to $7.1~{\rm MeV}$ for 
the various final states, is less than 
the natural width of the \chicz, but greater than the
natural width of the \chico\ 
and \chict.
The $\chi_{c}$
masses are fixed to their nominal values~\cite{PDG} during the fitting. 
In all cases, the 
reconstructed masses, when allowed to float, are consistent with 
the expected values.

\begin{figure}
\includegraphics*[width=6.0in]{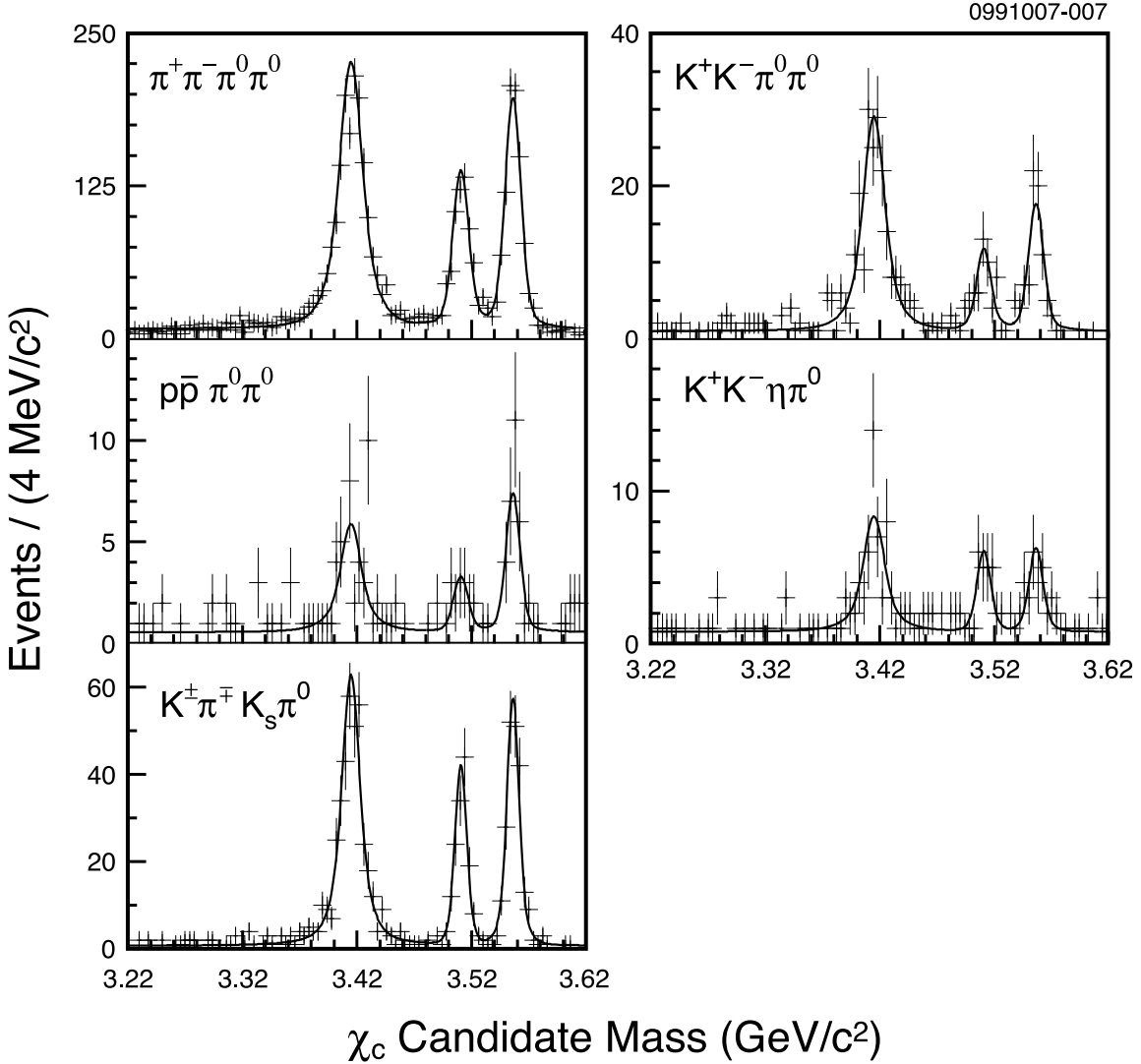}
\caption{The fitted distributions of the invariant mass of $\chi_c $ candidates in data. 
The fitting procedure is described in the text.} 
\label{fig:DataParents}
\end{figure}

Clean signals of \chicz, \chico\,
and \chict\ are found in most of the modes studied as seen in
Figure~\ref{fig:DataParents}. The signal yields and efficiencies, $\varepsilon$,
are listed in Table~\ref{Table:table_YandE_paper_NR} for each final state, calculated
assuming four-body phase space. 
The errors on the efficiencies due to limited simulation
statistics are considered with other
systematic uncertainties listed in Table~\ref{Table:table_sys}. The
yields and efficiencies in Table~\ref{Table:table_YandE_paper_NR} are
used to calculate the final branching fractions for all
modes, except \pipipiopio\ and \KpiKopio, which we proceed to study in more detail. 


\begin{table}[htb]
 \caption{Yields ($N$) and efficiencies $\varepsilon $  for
 four-hadron final states. Efficiencies were estimated using a  
Monte Carlo sample generated according to four-body phase space
as described in the text.}
 \begin{tabular}{l|cc|cc|cc}
 \hline
 \hline
  Mode  
 &\multicolumn{2}{c|}{$\chi_{c0}$}
 &\multicolumn{2}{c|}{$\chi_{c1}$} 
 &\multicolumn{2}{c}{$\chi_{c2}$} \\
 \hline
&$N$ &$\varepsilon$(\%) 
&$N$ &$\varepsilon$(\%) 
&$N$ &$\varepsilon$(\%) \\
 \hline
 \pipipiopio
  &1751.4$\pm$51.3 &17.6   
  &604.7$\pm$28.5  &17.5    
  &903.5$\pm$32.6  &17.6   \\
 \KKpiopio
 &213.5$\pm$16.8 &12.7    
 &45.1$\pm$8.5  &13.1    
 &76.9$\pm$9.7  &12.6 \\
 \pppiopio
 &39.5$\pm$8.5  &12.8 
 &11.5$\pm$4.2  &13.9   
 &29.2$\pm$5.9  &13.3 \\
 \KKetapio
 &56.4$\pm$9.2 &6.19    
 &21.0$\pm$5.7 &6.17   
 &22.9$\pm$6.1 &6.07 \\
 \KpmpimpKspio
 &401.7$\pm$22.4 &10.3    
 &141.3$\pm$13.3 &11.2    
 &211.6$\pm$15.4 &10.3 \\
 \hline
  \hline
 \end{tabular}
 \label{Table:table_YandE_paper_NR}
 \end{table}

\begin{figure}
\includegraphics*[width=5.5in]{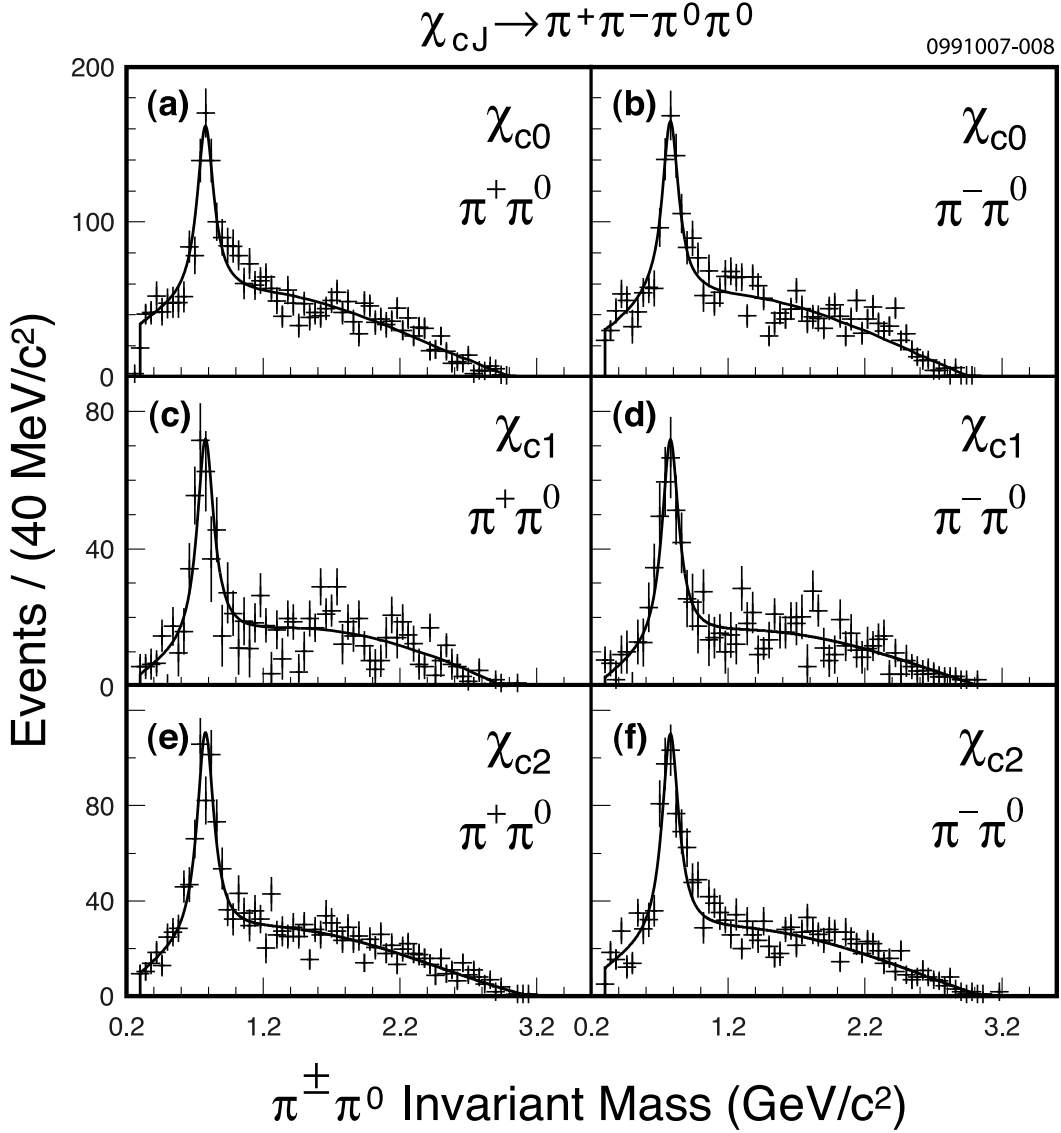}
\caption{Intermediate states for the decay $\chi_{cJ} \to
$ \pipipiopio. Invariant mass combinations of 
(a) $\pi^{+}\pi^{0} $ for $J=0 $, (b) $\pi^{-}\pi^{0} $ for $J=0 $, 
(c) $\pi^{+}\pi^{0} $ for $J=1 $, (d) $\pi^{-}\pi^{0} $ for $J=1 $, 
(e) $\pi^{+}\pi^{0} $ for $J=2 $, and (f) $\pi^{-}\pi^{0} $ for $J=2 $ are
shown after sideband subtraction. All plots show significant evidence for
production of $\rho^{\pm}$.}
\label{fig:DatasubM1}
\end{figure}

\begin{figure}
\includegraphics*[width=5.5in]{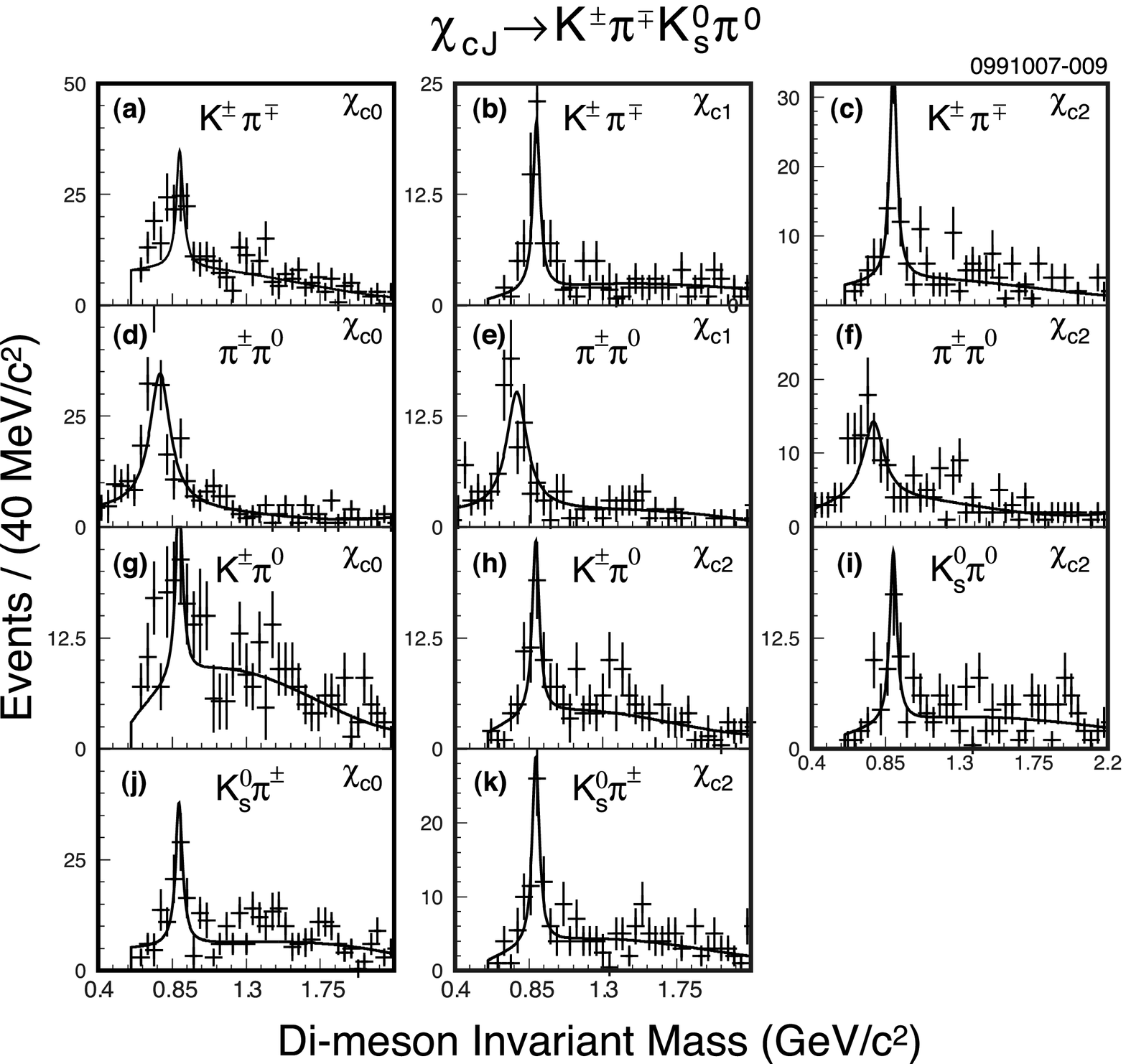}
\caption{Figure shows intermediate states for the decay $\chi_{cJ} \to
$ \KpmpimpKspio. Invariant mass combinations of 
(a) $K^{\pm}\pi^{\mp} $ for $J=0$,
(b) $K^{\pm}\pi^{\mp} $ for $J=1$,
(c) $K^{\pm}\pi^{\mp} $ for $J=2$, 
(d) $\pi^{\pm}\pi^{0} $ for $J=0$,
(e) $\pi^{\pm}\pi^{0} $ for $J=1$,
(f) $\pi^{\pm}\pi^{0} $ for $J=2$,
(g) $K^{\pm}\pi^{0} $ for $J=0$,
(h) $K^{\pm}\pi^{0} $ for $J=2$,
(i) $K_{S}^{0}\pi^{0} $ for $J=2$
(j) $\pi^{\pm}K_{S}^{0} $ for $J=0$, and
(k) $\pi^{\pm}K_{S}^{0} $ for $J=2$
are shown after sideband subtraction. Strong signals for
$\rho^{\pm} $ or $K^{*} $ production are visible.} 
\label{fig:DatasubM10}
\end{figure}

We can further investigate how the decays proceed
by searching for substructure in the four-hadron final states.
Substructure can also affect the
detection efficiency.
Here we restrict ourselves to
looking for the gross features of substructure by plotting
the invariant mass of di-hadron combinations
in the $\chi_c$ signal regions, after subtracting the 
yields from $\chi_c$ sidebands.
The signal regions for the invariant mass distributions of \chicz,
\chico, and \chict\ candidates, are 3.370-3.470~\gevcsq,
3.490-3.535~\gevcsq\ and 3.535-3.590~\gevcsq, respectively. One-sided
invariant mass distribution sidebands of 3.300-3.360~\gevcsq,
3.470-3.490~\gevcsq\ and 3.590-3.625~\gevcsq, respectively, are
subtracted. 
We claim and show signals only for those cases where we observe a
signal for resonant substructure with
significance of~$\geq $~4~$\sigma$; these are all in the
four body-modes 
\pipipiopio\ and \KpmpimpKopio.
After identifying the intermediate states, we
fit the distributions with Breit-Wigner functions for each of these
intermediate states using a fixed mass and width as listed in~\cite{PDG}, 
and add an additional third order polynomial function to
describe the background. We obtain the efficiencies for detecting the
intermediate states, $\varepsilon^{\prime}$, by fitting the resonant
signals using the procedure described above, 
in samples of events that simulated the decay as proceeding entirely through
the respective intermediate states.
%
Figures~\ref{fig:DatasubM1} and~\ref{fig:DatasubM10} show two-body
intermediate states in \pipipiopio\ and \KpmpimpKspio\
final states, respectively, for each of the \chic\
states. In the former case there are two combinations entering the 
plot for each $\chi_c$ candidate.
We find clear signals for the $\rho $ and $K^{*} $ mesons.
The yields and efficiencies, $\varepsilon^{\prime} $, of final states
including these resonances are listed in 
Table~\ref{Table:table_YandE_paper_R} and are used as inputs to
the final branching fraction calculations. 

Several sources of systematic uncertainty in the branching fractions
are listed in Table~\ref{Table:table_sys}. 
These include uncertainties in the number of \Psip particles, determined according to the method described in~\cite{psi2stogamchicj},
tracking and particle identification efficiencies associated
with charged particles, reconstruction efficiencies due to simulation
statistics, and trigger simulation. 
By comparing the effect of varying
the requirement on the  $\chi^2 $ of the constrained four-momentum fit
 in data and in simulated
events, we estimate a $\pm$4.0\% systematic error in modelling the
$\chi^2 $ distribution.
The uncertainty in the efficiency of the
transition photon reconstruction for each of the final states
is taken to be 2\%. In
addition, the systematic uncertainties for \pio\ and  $\eta $ efficiencies
are 4\% for each \pio\ or $\eta $ meson in the final state. 
\Ks\ reconstruction
introduces an additional uncertainty of 2\% in one of the modes. \
The robustness of the fitting procedure was checked by systematically
re-fitting the $\chi_{c} $ invariant mass plots using 1~$\sigma $
variations of masses, widths, and resolutions.
The uncertainties estimated from this study
vary between 0.2\% and 0.7\%.
To account for the systematic error arising from deviations in the angular
distributions of final states due to the presence of intermediate
resonances, a sample of events that simulated the decay as proceeding through
the respective intermediate states, was generated for the substructure modes:
\fonpiopio, \ftpiopio, \rhopmpimppio\ and \fonpippim\ for the
non-resonant mode \pipipiopio; \KstarpmKmppio\ and \fonKpKm\ for the
non-resonant mode \KKpiopio; 
$K^{*0}K_{S}^{0}\pi^{0}$, $K^{*\pm}K_{S}^{0}\pi^{\mp}$,
\KstarpmKmppio, $\rho^{\pm}K^{\mp}K_{S}^{0}$, \KstaroKpmpimp\ and
$K_{1}(1270)K_{S}^{0} $ for the
non-resonant mode \KpmpimpKspio . 
We obtained the efficiencies of both
the resonant and non-resonant modes by fitting the $\chi_{c} $
signals obtained from MC events with and without substructure. Based on the
differences in efficiencies between final states with and without
intermediate resonances, and assuming that there can be no additional
unobserved resonances that can be more than 50\% of the signal, we estimate the systematic error to be
between 3.5\% and 7.4\% for
the modes we studied. For the remaining modes, we conservatively assign a 
7.5\% systematic error based on an assumption that up to 75\% of our events 
may contain substructure, 
and that the difference in efficiency of these resonant
events to the non-resonant is no more than 10\%.
 \begin{table}[htb]
 \caption{Yields and efficiencies (in \%) for substructure modes. 
$\varepsilon $ represents the efficiency obtained by 
 fitting the \chic\ signals using the same procedure as that used for the
 non-resonant four-hadron modes, in substructure simulated samples. Efficiency
 $\varepsilon^{\prime} $ was obtained by fitting the intermediate peaks after
 applying the sideband subtraction procedure described in the text. $N^{\prime}$
describes the yield corresponding to this efficiency $\varepsilon^{\prime}$.}
 \begin{tabular}{l|ccc|ccc|ccc}
 \hline
 \hline
 Mode  
 &\multicolumn{3}{c|}{ $\chi_{c0}$}
 &\multicolumn{3}{c|}{ $\chi_{c1}$} 
 &\multicolumn{3}{c}{ $\chi_{c2}$} \\
 \hline
 &$N^{\prime} $&$\varepsilon $ &$\varepsilon^{\prime} $  
 &$N^{\prime} $&$\varepsilon $ &$\varepsilon^{\prime} $ 
 &$N^{\prime} $&$\varepsilon $ &$\varepsilon^{\prime} $  \\
 \hline
 \rhoppimpio
 &661.4$\pm$56.8 &17.4 &15.7  
 &355.7$\pm$42.1 &16.9 &16.3 
 &519.3$\pm$38.7 &16.8 &16.1 \\
\rhompippio
 &697.1$\pm$56.6 &17.4 &15.7  
 &356.6$\pm$41.0 &16.9 &16.3 
 &512.6$\pm$40.0 &16.8 &16.1 \\
 \hline
 $K^{*0}K_{S}^{0}\pi^{0}$, \KstarotoKpmpimp   
 &52.5$\pm$12.1 &11.0 &9.94 
 &37.9$\pm$8.5  &11.1 &10.9
 &63.0$\pm$10.5 &11.5 &11.2 \\
 \KstaroKpmpimp, $K^{*0} \to K_{S}^{0}\pi^{0}$ 
 & - & &
 & - & &
 &38.7$\pm$9.0  &9.50 &9.14 \\
 \KstarpmKmppio, $K^{*\pm} \to \pi^{\pm} K_{S}^{0}$   
 &64.1$\pm$12.8 &10.3 &9.30
 &- & & 
 &51.1$\pm$9.8 &9.66 &9.40 \\
 $K^{*\pm}\pi^{\mp}K_{S}^{0}$, \KstarpmtoKpmpio  
 &42.5$\pm$10.0 &10.4 &9.44 
 & - & &
 &39.3$\pm$8.7 &9.44 &9.10 \\
 $\rho^{\pm}K^{\mp}K_{S}^{0}$  
 &179.7$\pm$22.7 &11.0 &9.92   
 &79.5$\pm$16.9  &10.8 &10.6  
 &62.9$\pm$15.9  &10.8 &10.5 \\
 \hline
 \hline
 \end{tabular}
 \label{Table:table_YandE_paper_R}
 \end{table}

 \begin{table}[htb]
 \caption{Systematic uncertainties (fractional errors in \%). The overall systematic uncertainty
is obtained by adding the individual contributions in quadrature, with the exception
of the photon simulation uncertainty which adds linearly with the $\pi^0$ and $\eta$
reconstruction uncertainties.}
 \begin{tabular}{cccccc}
 \hline
 \hline
 Source &\ \pipipiopio\ &\ \KKpiopio\ &\ \pppiopio \ &\ \KKetapio\ &\ \KpmpimpKspio\ \\
\hline
 $N_{\psi(2S)} $ &3.0 &3.0 &3.0 &3.0 &3.0 \\
 Tracking efficiency &1.4 &1.4 &1.4 &1.4 &2.8 \\ 
 Particle identification &0.6 &2.6 &2.6 &2.6 &2.2\\
 Simulation statistics &1.9 &2.3 &2.3 &2.8 &4.5 \\
 Trigger efficiency &1.0 &1.0 &1.0 &1.0 &1.0\\
 Kinematic constraint cut &4.0 &4.0 &4.0 &4.0 &4.0 \\
 Transition $\gamma $ simulation &2.0 &2.0 &2.0 &2.0 &2.0 \\
 $\pi^0,\eta \to \gamma \gamma $ reconstruction &8.0 &8.0 &8.0 &8.0 &4.0 \\
 \Ks\ Vertexing &0.0 &0.0 &0.0 &0.0 &2.0 \\
 Fitting procedure &0.5 &0.6 &0.7 &0.7 &0.2 \\
 Model dependence &3.5 &7.4 &7.5 &7.5 &6.0 \\
\hline
  {\bf Overall} &{\bf 12.0} &{\bf 14.0} &{\bf 14.0} &{\bf 14.1} &{\bf 11.6 } \\ 	
\hline
\hline	
 \end{tabular}
 \label{Table:table_sys}
 \end{table}

Systematic uncertainties for the intermediate state resonance modes
include common sources to their non-resonant
final states.  
Uncertainties were added in quadrature, 
and those due to the branching fractions of \Psip $\to \gamma \chi_{c}$~\cite{psi2stogamchicj} are quoted
separately. We use the values of ${\cal B}$(\Psip $\to
\gamma \chi_{c}$) in ~\cite{psi2stogamchicj}, as
those measurements use a subset of our data and similar analysis criteria,
thereby enhancing the cancellation of systematic errors.
For all final states except $\pi^+\pi^-\pi^0\pi^0\ $ and
$K^{\pm}\pi^{\mp}K_s\pi^0$, we convert the yields in
Tables~\ref{Table:table_YandE_paper_NR} and~\ref{Table:table_YandE_paper_R} to branching fractions using: 

\begin{equation}
        {\cal{B}} ( \chi_{cJ} \to i )  =
        \frac{N_{i}} {N_{\psi(2S)} \cdot \varepsilon_{i} 
	\cdot {\cal B}(\psi(2S) \to \gamma \chi_{cJ})} 
\label{eqn:BF formula}
\end{equation}
where $N_i$ is the yield; $N_{\psi(2S)}$ is the  number of
$\psi(2S) = $ 3.08 $\times 10^6 $; $\varepsilon$ is the same
$\varepsilon $ listed in Table~\ref{Table:table_YandE_paper_NR} for four-hadron modes,
or $\varepsilon^{\prime} $ listed in
Table~\ref{Table:table_YandE_paper_R} for substructure modes,
 and $i $ represents a particular decay mode.


To calculate the branching fractions for the inclusive four-hadron final
states for \pipipiopio\ and \KpmpimpKopio\ 
we use a modified procedure. Since the \rhopmpimppio\
clearly dominates the four-hadron final state yields
for the \pipipiopio\ mode (Tables~\ref{Table:table_YandE_paper_NR} 
and~\ref{Table:table_YandE_paper_R}), we
use the efficiency $\varepsilon $ of the
\rhopmpimppio\ sub-mode listed in Table~\ref{Table:table_YandE_paper_R} and 
Equation~\ref{eqn:BF formula} to determine the \pipipiopio\ 
branching fraction. The efficiencies $\varepsilon $ were 
obtained by fitting the $\chi_{cJ} $ signals in substructure
simulations using the same fitting procedure as that used for the
four-hadron signal simulations\footnote{The efficiency
$\varepsilon^{\prime} $ is lower than $\varepsilon $
since the sideband subtraction procedure used for
obtaining the efficiency $\varepsilon^{\prime} $ results in some loss of
efficiency.}.
To calculate the \KpmpimpKopio\ branching fraction, we modify the
procedure by taking into account that this channel has many
intermediate resonances (Table~\ref{Table:table_YandE_paper_R}). 
We replace the ratio $N_{i}/e_{i} $ in Equation~\ref{eqn:BF
formula} with an efficiency corrected yield ($Y_{K^{\pm}\pi^{\mp}K^{0}\pi^{0}}$) by adding the individual
efficiency corrected contributions of all resonant and
non-resonant channels computed as:
\begin{equation}
Y_{K^{\pm}\pi^{\mp}K^{0}\pi^{0}} =
\frac{N_{K^{\pm}\pi^{\mp}K^{0}\pi^{0}}}{\varepsilon_{K^{\pm}\pi^{\mp}K^{0}\pi^{0}}}
+ \sum_{k=1}^{5} \frac{N_{k}}{\varepsilon_{k}^{\prime}} \cdot \left(1-\frac{\varepsilon_{k}}
{\varepsilon_{K^{\pm}\pi^{\mp}K^{0}\pi^{0}}} \right)
\label{eqn:kpikopio Yec}
\end{equation}
where $k $ runs over the substructure modes we consider,
$K^{*0}K_{S}^{0}\pi^{0}$, \KstaroKpmpimp, \KstarpmKmppio,
$K^{*\pm}\pi^{\mp}K_{S}^{0}$, and $\rho^{\pm}K^{\mp}K_{S}^{0}$.

The branching fractions  obtained and
their uncertainties are summarized in Table~\ref{Table:tableofBFsall}.
Where we do not find evidence of a signal, we present a 90\% C.L.  
upper limit by determining the value that includes 90\% of the  
probability density function (p.d.f) obtained by convolving the p.d.f  
for the branching fraction with a Gaussian systematic error.

  
The results of the branching fractions of the four-hadron modes
studied in this analysis include both 
resonant and non-resonant
contributions. Our three-hadron intermediate resonance 
branching fractions are inclusive and may contain additional resonant substructure 
that we do not explicitly measure; therefore, the branching fraction measurements
cannot be trivially related to the amplitude for the specific three-body non-resonant
decay.
We assume variations of the efficiency due to this additional substructure produce changes in the 
branching fractions within the model dependence systematic error.

We compare the branching fractions $\chi_{cJ} \to $ \rhopmpimppio\ for
$J=0,1,2 $ (Table~\ref{Table:tableofBFsall}) with measurements of
$\chi_{cJ} \to $ \rhozpippim~\cite{PDG}, and observe that our results are consistent
with a ratio equal to unity as expected from isospin conservation.
 
Furthermore, we also find our measurement of $\cal{B}$(\chict\ $\to $ \KstaroKpmpimp) 
(Table~\ref{table:table_br_isospin}) to be consistent within 
experimental errors with $\cal{B}$(\chict\ $\to $ \KstaroKpmpimp) of~\cite{PDG}.
Our results are consistent with
the expected isospin-related prediction that 
$\chi_{c} \to K^{*}K\pi $ where $\chi_{c} \to $ \KstaroKopio\ and
$\chi_{c} \to $ \KstarpmKmppio\, have equal partial
widths.
Moreover, our measurements for $\cal{B}$(\chict\ $\to $ \KstaroKpmpimp), 
$\cal{B}$(\chicz\ $\to $ \KstarpmpimpKo), and 
$\cal{B}$(\chict\ $\to $\KstarpmpimpKo) (Table ~\ref{table:table_br_isospin}) are in good agreement with the isospin expectation of 


\begin{equation}
{\cal B}(\chi_{c} \to K^{*0}K^{0}\pi^{0}) : {\cal B}(\chi_{c} \to
K^{*0}K^{\pm}\pi^{\mp})=1:2
\end{equation}
and
\begin{equation}
{\cal B}(\chi_{c} \to K^{*0}K^{0}\pi^{0}) : {\cal B}(\chi_{c} \to K^{*\pm}\pi^{\mp}K^{0})=1:2. 
\end{equation}


\begin{table}[htb]
 \caption{Branching fractions ${\mathcal B}$ with statistical and systematic
 uncertainties are shown. The symbol ``$\times$'' indicates product of
 branching fractions. The third error in each case is
 due to the \Psip $\to \gamma \chi_{c} $ branching fractions. Upper limits
 shown are at the 90\% C.L and include all the systematic errors. The first line
of each set of measurements includes the contributions from the substructure listed below.
}

 \begin{tabular}{lccc}
 \hline
 \hline
 Mode  & { $\chi_{c0}$ }& {$\chi_{c1}$} & { $\chi_{c2}$} \\
 \hline
 & $\mathcal{B}$~(\%) & $\mathcal{B}$~(\%) & $\mathcal{B}$~(\%) \\
 \hline
 \pipipiopio
 &    3.54$\pm$0.10$\pm$0.43$\pm$0.18
 &    1.28$\pm$0.06$\pm$0.15$\pm$0.08
 &    1.87$\pm$0.07$\pm$0.22$\pm$0.13 \\
 &    
 &    
 &      \\
 \ \rhoppimpio
 &1.48$\pm$0.13$\pm$0.18$\pm$0.08
 &0.78$\pm$0.09$\pm$0.09$\pm$0.05
 &1.12$\pm$0.08$\pm$0.13$\pm$0.08 \\
 \ \rhompippio
 &1.56$\pm$0.13$\pm$0.19$\pm$0.08
 &0.78$\pm$0.09$\pm$0.09$\pm$0.05
 &1.11$\pm$0.09$\pm$0.13$\pm$0.08 \\
 \hline
 \KKpiopio
 &    0.59$\pm$0.05$\pm$0.08$\pm$0.03
 &    0.12$\pm$0.02$\pm$0.02$\pm$0.01
 &    0.21$\pm$0.03$\pm$0.03$\pm$0.01 \\
 \hline
 \pppiopio
 &    0.11$\pm$0.02$\pm$0.02$\pm$0.01
 &    $< 0.05 $
 &    0.08$\pm$0.02$\pm$0.01$\pm$0.01 \\
 \hline
 \KKetapio
 &    0.32$\pm$0.05$\pm$0.05$\pm$0.02
 &    0.12$\pm$0.03$\pm$0.02$\pm$0.01
 &    0.13$\pm$0.04$\pm$0.02$\pm$0.01 \\
 \hline
 \KpmpimpKopio
 &    2.64$\pm$0.15$\pm$0.31$\pm$0.14
 &    0.92$\pm$0.09$\pm$0.11$\pm$0.06
 &    1.41$\pm$0.11$\pm$0.16$\pm$0.10 \\
 &    
 &    
 &      \\
 \ \KstaroKopio $\times$ \KstarotoKpmpimp
  &0.37$\pm$0.09$\pm$0.04$\pm$0.02
 &0.25$\pm$0.06$\pm$0.03$\pm$0.02
 &0.39$\pm$0.07$\pm$0.05$\pm$0.03 \\
 \ \KstaroKpmpimp $\times$ \KstarotoKopio
 & -
 & -
 &0.30$\pm$0.07$\pm$0.04$\pm$0.02 \\
 \ \KstarpmKmppio $\times$ \KstarpmtopipmKo
 &0.49$\pm$0.10$\pm$0.06$\pm$0.03
 & -
 &0.38$\pm$0.07$\pm$0.04$\pm$0.03 \\ 
 \ \KstarpmpimpKo $\times$ \KstarpmtoKpmpio 
 &0.32$\pm$0.07$\pm$0.04$\pm$0.02
 & -
 &0.30$\pm$0.07$\pm$0.04$\pm$0.02 \\
 \ \rhopmKmpKo
 &1.28$\pm$0.16$\pm$0.15$\pm$0.07
  &0.54$\pm$0.11$\pm$0.06$\pm$0.03
  &0.42$\pm$0.11$\pm$0.05$\pm$0.03 \\
 \hline
 \hline
 \end{tabular}
 \label{Table:tableofBFsall}
 \end{table}


 \begin{table}[htb]
 \caption{Branching fractions and total error measurements for the isospin-related $K^{*}K\pi$ intermediate modes.}
 \begin{tabular}{lccc}
 \hline
 \hline
 Mode  &{ $\chi_{c0}$ }& {$\chi_{c1}$} &{ $\chi_{c2}$} \\
 \hline
 &$\mathcal{B}$~(\%) & $\mathcal{B}$~(\%) & $\mathcal{B}$~(\%)\\
 \hline
 \KstaroKopio &0.56$\pm$0.15 &0.38$\pm$0.11 &0.59$\pm$0.14 \\
 \KstaroKpmpimp &- &-  &0.90$\pm$0.25 \\
 \KstarpmKmppio &0.74$\pm$0.18 &-  &0.57$\pm$0.13 \\
 \KstarpmpimpKo &0.96$\pm$0.25 &- &0.90$\pm$0.25\\
 \hline
 \hline
 \end{tabular}
 \label{table:table_br_isospin}
 \end{table}

In summary, the
branching fractions for $\chi_{cJ} \to $ \pipipiopio, $\chi_{cJ} \to $ \KKpiopio, $\chi_{cJ} \to $ \KKetapio, and $\chi_{cJ} \to $ \KpmpimpKopio\ for $J = 0,1,2 $ and, $\chi_{cJ} \to $ \pppiopio\ for $J = 0,2 $ 
are measured for the first time.
For the mode $\chi_{c1} \to $ \pppiopio\ upper limits at
90\% C.L. are presented. We also measure for the first time the
partial branching fractions of 
$\chi_{cJ} \to $\rhompippio, $\chi_{cJ} \to $ \KstaroKopio, and $\chi_{cJ} \to $
\rhopmKmpKo\ for $J = 0,1,2 $; $\chi_{c2} \to $ \KstaroKpmpimp;
$\chi_{cJ} \to $ \KstarpmKmppio and $\chi_{cJ} \to $ \KstarpmpimpKo\
for $J = 0,2 $. 
These four-hadron final states account for up to 8\% of the hadronic width of the \chic\ states.

These measurements improve our existing knowledge of the exclusive
multi-body decay modes of the \chic\ states and provide insight into
their decay
mechanisms. 
%
The four-hadron final states \pipipiopio\ and 
\KpmpimpKopio\
contain a 
rich substructure of 
intermediate resonances. 
These interesting results form a basis from which higher statistics
studies of hadronic \chic\ states and their substructure may follow.

We gratefully acknowledge the effort of the CESR staff
in providing us with excellent luminosity and running conditions.
D.~Cronin-Hennessy and A.~Ryd thank the A.P.~Sloan Foundation.
This work was supported by the National Science Foundation,
the U.S. Department of Energy, and
the Natural Sciences and Engineering Research Council of Canada.

\end{document}